\documentclass[usenatbib]{mnras}
\usepackage{epsfig,booktabs,caption,times,graphicx,amsmath}
\usepackage[flushleft]{threeparttable}

\def\Hy@FixNotFirstPage{%
      \gdef\Hy@FixNotFirstPage{%
            \setbox\AtBeginShipoutBox=\hbox{%
                  \copy\AtBeginShipoutBox
            }%
      }%
}
\AtBeginShipout{\Hy@FixNotFirstPage}
 
\def\I{\,\textsc{i}}
\def\II{\,\textsc{ii}}

\def\hst{{\it HST}}

\def\ubvgri{{$uBV\!gri$}}

\title[Limits on the progenitor of the SN~Ia 2017ejb]{X-ray Limits on the Progenitor System of the Type Ia Supernova 2017ejb}

\author[Kilpatrick et~al.]{Charles D. Kilpatrick$^1$\thanks{Email:
    cdkilpat@ucsc.edu}, David~A.~Coulter$^1$, Georgios~Dimitriadis$^1$, Ryan~J.~Foley$^1$,
    \newauthor David~O.~Jones$^1$, Yen-Chen~Pan$^1$, Anthony~L.~Piro$^2$, Armin~Rest$^{3,4}$, C\'{e}sar Rojas-Bravo$^1$ \\
      $^1$Department of Astronomy and Astrophysics, University of California, Santa Cruz, CA 95064, USA\\
      $^2$The Observatories of the Carnegie Institution for Science, 813 Santa Barbara St., Pasadena, CA 91101, USA\\
      $^3$Space Telescope Science Institute, 3700 San Martin Drive, Baltimore, MD 21218, USA\\
      $^4$Department of Physics and Astronomy, The Johns Hopkins University, 3400 North Charles Street, Baltimore, MD 21218, USA}

\begin{document}
\date{Accepted 0000, Received 0000, in original form 0000}
\pagerange{\pageref{firstpage}--\pageref{lastpage}} \pubyear{2018}
\maketitle
\label{firstpage}

\begin{abstract}

We present deep X-ray limits on the presence of a pre-explosion counterpart to the low-luminosity Type Ia supernova (SN~Ia) 2017ejb.  SN~2017ejb was discovered in NGC~4696, a well-studied elliptical galaxy in the Centaurus cluster with $894$~ks of {\it Chandra} imaging between $14$ and $3$~years before SN~2017ejb was discovered. Using post-explosion photometry and spectroscopy of SN~2017ejb, we demonstrate that SN~2017ejb is most consistent with low-luminosity SNe~Ia such as SN~1986G and SN~1991bg. Analyzing the location of SN~2017ejb in pre-explosion images, we do not detect a pre-explosion X-ray source.  We use these data to place upper limits on the presence of any unobscured supersoft X-ray source (SSS). SSS systems are known to consist of white dwarfs accreting from a non-degenerate companion star.  We rule out any source similar to known SSS systems with $kT_{\rm eff} > 85$~eV and $L_{\rm bol} > 4\times10^{38}~\text{erg s}^{-1}$ as well as models of stably-accreting Chandrasekhar-mass WDs with accretion rates $\dot{M}>3\times10^{-8}~M_{\odot}~\text{yr}^{-1}$.  These findings suggest that low-luminosity SNe~Ia similar to SN~2017ejb explode from WDs that are low-mass, have low pre-explosion accretion rates, or accrete very soon before explosion.  Based on the limits from SN~2017ejb and other nearby SNe~Ia, we infer that $<$47\% of SNe~Ia explode in stably-accreting Chandrasekhar-mass SSS systems.
  
\end{abstract}

\begin{keywords}
      supernovae: general --- supernovae: individual (SN~2017ejb) --- X-rays: general
\end{keywords}

\section{Introduction}\label{s:intro}

Type Ia supernovae (SNe Ia) are a homogeneous class of SNe defined by a lack of hydrogen and helium in their spectra but with strong silicon absorption \citep[for a review see, e.g.][]{filippenko97}. For over 50 years, the leading progenitor model for SNe~Ia has been a white dwarf (WD) that undergoes a thermonuclear explosion \citep{hf60,finzi+67,hansen+69}. The pathway by which these WDs ignite is less certain. Potential models include the merger of two carbon/oxygen WDs \citep{iben+84,webbink+84}, accretion and detonation of a helium shell on a sub-Chandrasekhar WD \citep{taam+80,shen+14}, direct collision of two unbound WDs in dense stellar systems \citep{rosswog+09,raskin+10,thompson+11,kushnir+13}, or steady accretion leading to a Chandrasekhar-mass explosion \citep{whelan+73,nomoto+82}. Even among these general classes of explosion scenarios there are important differences, such as whether ignition in the merger case is triggered by unstable \citep{guillochon+10,dan+12,pakmor+12} or stable mass transfer \citep{fink+07,fink+10,shen+09}.

The fact that these models are all theoretically plausible and reproduce some of the observed characteristics of SNe~Ia may reflect SN~Ia diversity within the overall class.  Despite the homogeneity of SN~Ia spectroscopic features and light curve shapes, they span a range of luminosities \citep[from low-luminosity 1991bg-like SNe~Ia to high-luminosity 1991T-like and 2006gz-like SNe~Ia; e.g.,][]{phillips+99,ashall+16}, ejecta velocities \citep{foley+10,mandel+14}, abundance distributions in their outer ejecta layers \citep[][]{lentz+00,foley+16,cartier+16}, abundances inferred from nebular spectra \citep{mazzali+15}, and large-scale environments \citep{cooper+09,sullivan+10,pan+14}.  Detailed predictions for how these properties depend on explosion scenario are one of the most promising avenues for determining the true explosion pathway(s) \citep[][]{foley+12,maoz+14}.

In addition to understanding their explosion physics, the connection between progenitor channels and SN~Ia luminosity is of utmost importance for cosmology.  SN~Ia light curves are among the most reliable redshift-independent distance indicators out to high redshift \citep[][]{jones+13,rubin+17}, and they are the basis for the discovery of the accelerating expansion of the Universe \citep{riess+98,perlmutter+99}.  However, as we measure larger samples of SN~Ia light curves with increasing precision, it has become clear that a major limiting factor in using SNe~Ia to measure cosmological parameters is systematic uncertainty in how SN~Ia explosion properties affect their intrinsic colors and luminosity \citep[see analysis in][]{scolnic+17}.  A physically-motivated understanding of SN~Ia evolution at all wavelengths is essential before these systematic uncertainties can be thoroughly addressed and precision in cosmological parameters is significantly improved.  Fundamentally, this means isolating an explosion model and observables that break the degeneracies between SN~Ia light curve shape and intrinsic luminosity.

Various explosion models predict radically different pre-explosion states for SNe~Ia, including electromagnetic and gravitational signals that may be detectable from nearby systems.  Inspiraling binary WDs produce a background of gravitational wave emission (in the 0.1--1~mHz regime) that will be targeted and potentially resolvable by {\it LISA} \citep{edlund+05}.  Accreting WDs with non-degenerate companion stars produce thermal emission that peaks in the ultraviolet and X-ray \citep{distefano+09}.  Sufficiently massive and luminous WD companion stars may be directly observed in pre-explosion images of nearby SNe~Ia \citep[][]{maeda+14}. These signals have been explored for some nearby systems; for example, optical pre-explosion limits \citep[for SNe~2011fe and 2014J;][]{li+11,kelly+14} have ruled out $>5~M_{\odot}$ companions for two ``normal'' SNe~Ia \citep[i.e., similar to those used for cosmology in][]{riess+16}.  Many nearby galaxies are well-studied at X-ray energies with deep {\it Chandra} imaging, and SNe~2011fe and 2014J have deep limits on the presence of an accreting WD, also called a supersoft X-ray source (SSS), in a symbiotic binary or accreting from the wind of its companion star \citep{nielsen+12,nielsen+13,nielsen+14}.

In rare cases, deep pre-explosion imaging can serendipitously lead to interesting limits on SN~Ia progenitor systems, even for SNe that occur much more than $10$~Mpc away \citep[whereas, e.g., SNe~2011fe and 2014J were 7.2 and 3.5~Mpc away, respectively;][]{li+11,kelly+14}.  This was the case for SN~2012fr, whose host galaxy was observed by {\it Chandra} for a total of $\sim$300~ks, providing the third deepest limits on the presence of a SSS (after SNe~2011fe and 2014J) in spite of the fact that SN~2012fr is $21$~Mpc away \citep{nielsen+13}.  Similarly, the host galaxy of the low-luminosity SN~Iax 2012Z was observed by the {\it Hubble Space Telescope} (\hst) for $>100$~ks, and a blue source consistent with a non-degenerate helium companion star was identified despite the fact that it is $33$~Mpc away \citep[][]{mccully+14}.  Although nearby events in well-studied galaxies typically lead to deeper limits on the presence of a progenitor system, systematic follow up of all nearby SNe~Ia with pre-explosion imaging is essential to understand the progenitor population as a whole.

In this paper, we discuss SN~2017ejb, which was discovered in the elliptical galaxy NGC~4696 (the brightest galaxy in the Centaurus cluster) on 28.22 May 2017 by the D$<$40~Mpc (DLT40) survey \citep{atel10439}\footnote{SN~2017ejb is also called DLT17bk}.  Deep limits from 6~days before discovery suggest that SN~2017ejb was first observed within a few days of explosion.  Follow-up spectroscopy of SN~2017ejb on 29 May 2017 \citep{atel10437,atel10438} suggested that it was a 1991bg-like SN~Ia roughly one week before maximum light.

Here, we report pre-explosion {\it Chandra} and \hst\ imaging of the explosion site of SN~2017ejb as well as follow-up photometry and spectroscopy.  Our light curves and spectra indicate that SN~2017ejb is a peculiar SN~Ia with a low peak luminosity, lacks a secondary $i$-band maximum, and has strong carbon absorption at early times.  Overall, this source is most similar to low-luminosity SNe~Ia such as  SN~1986G and SN~1991bg.  We examine all pre-explosion data to look for an optical or X-ray counterpart to SN~2017ejb, but do not detect any sources.  The limiting X-ray flux rules out the presence of any SSS similar to known systems with bolometric luminosity $>4\times10^{38}~\text{erg s}^{-1}$ or effective temperature $>85$~eV.  These limits rule out much of the temperature-luminosity space for SSS systems in nearby galaxies as well as models of stably-accreting Chandrasekhar-mass WDs with accretion rates $\dot{M}>3\times10^{-8}~M_{\odot}~\text{yr}^{-1}$.

Throughout this paper, we assume a Milky Way reddening to NGC~4696 of $E(B-V)=0.098$~mag \citep{schlafly+11} and a distance to the Centaurus cluster of $d=41.3\pm2.1$~Mpc \citep[$\mu=33.08\pm0.11$~mag;][]{mieske+03}.

\section{Observations}

\subsection{Archival Data}\label{s:archival}

\subsubsection{{\it Chandra}}

We searched for pre-explosion observations of NGC~4696 from the {\it Chandra} Data Archive.  We found data consisting of 17 epochs of Advanced CCD Imaging Spectrometer (ACIS) images and totaling $\sim$894~ks of effective exposure time.  These data were obtained between 22 May 2000 and 5 Jun. 2014.  We list all {\it Chandra} observations in \autoref{tab:chandra}.

\begin{table}
\begin{center}\begin{minipage}{3.3in}
      \caption{{\it Chandra}/ACIS Data of NGC~4696}\small
\begin{tabular}{@{}cccc}\hline\hline
 {\it Chandra} & Epoch & Exposure & Pointing Center \\
 Observation   & (start date)&   (ks)        &  ($\alpha$,$\delta$) (J2000.0) \\ \hline
      504      & $-$6215.20  & 31.75           & 12:48:48.70,$-$41:18:44.00 \\
      505      & $-$6198.22  & 9.96            & 12:48:48.70,$-$41:18:44.00 \\
      1560     & $-$6248.53  & 84.75           & 12:48:49.40,$-$41:18:40.50 \\
      4190     & $-$5153.68  & 34.27           & 12:49:05.00,$-$41:16:17.00 \\
      4191     & $-$5153.26  & 34.02           & 12:48:41.00,$-$41:22:36.00 \\
      4954     & $-$4804.64  & 89.05           & 12:48:48.90,$-$41:18:44.40 \\
      4955     & $-$4803.58  & 44.68           & 12:48:48.90,$-$41:18:44.40 \\
      5310     & $-$4802.04  & 49.33           & 12:48:48.90,$-$41:18:44.40 \\
      8179     & $-$3716.64  & 29.79           & 12:50:03.90,$-$41:22:57.00 \\
      16608    & $-$1146.49  & 34.11           & 12:48:48.90,$-$41:18:43.80 \\
      16224    & $-$1144.85  & 42.29           & 12:48:48.90,$-$41:18:43.80 \\
      16607    & $-$1142.20  & 45.67           & 12:48:48.90,$-$41:18:43.80 \\
      16625    & $-$1127.24  & 30.10           & 12:48:48.90,$-$41:18:43.80 \\
      16610    & $-$1126.34  & 17.34           & 12:48:48.90,$-$41:18:43.80 \\
      16609    & $-$1119.92  & 82.33           & 12:48:48.90,$-$41:18:43.80 \\
      16223    & $-$1097.25  & 178.97          & 12:48:48.90,$-$41:18:43.80 \\
      16534    & $-$1087.85  & 55.44           & 12:48:48.90,$-$41:18:43.80 \\ \hline
\end{tabular}\label{tab:chandra}
\begin{tablenotes}
      \small
\item {\bf Note.} Epoch is in days relative to discovery on 28.22 May 2017.
\end{tablenotes}
\end{minipage}\end{center}
\end{table}

Using the {\it Chandra} Interactive Analysis of Observations (\textsc{ciao}) software package \citep{fruscione+99}, we merged all of these data into a single event map. We note that SSS emission is negligible above 1~keV \citep[1.2~nm; see][]{distefano+04,ness+13}, and so following similar procedures in \citet{nielsen+11}, we limited our analysis to events in the 0.3--1.0~keV soft band of {\it Chandra}/ACIS. We used \textsc{ciao}/{\tt merge\_obs} to construct event and exposure maps centered around the location of SN~2017ejb as reported in \citet{atel10439}.

\begin{figure*}
      \setlength{\fboxsep}{-0.25pt}
      \setlength{\fboxrule}{1.0pt}
      \begin{center}\begin{minipage}{6.8in}
      \fbox{\includegraphics[width=0.321\textwidth]{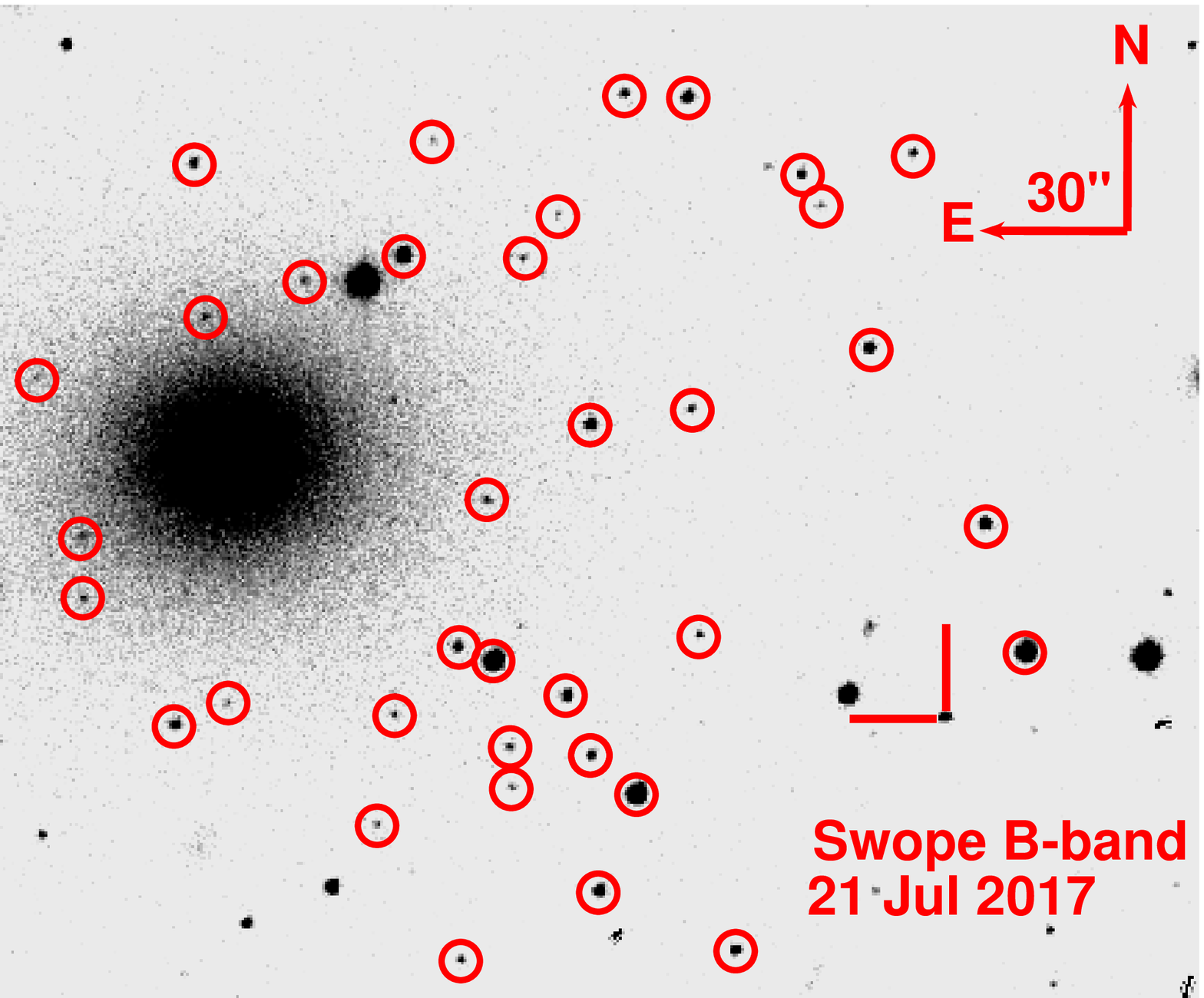}}
      \fbox{\includegraphics[width=0.321\textwidth]{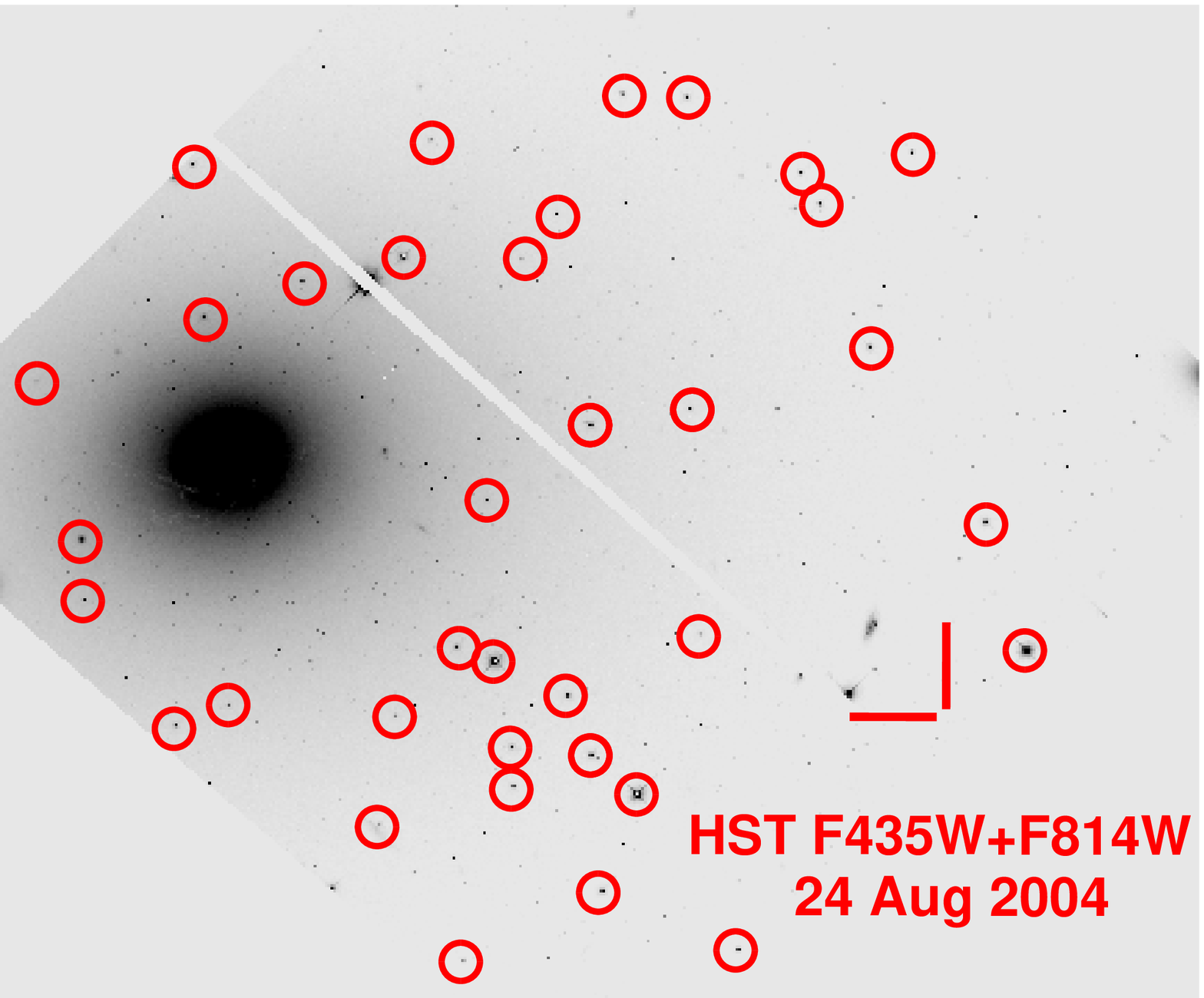}}
      \fbox{\includegraphics[width=0.321\textwidth]{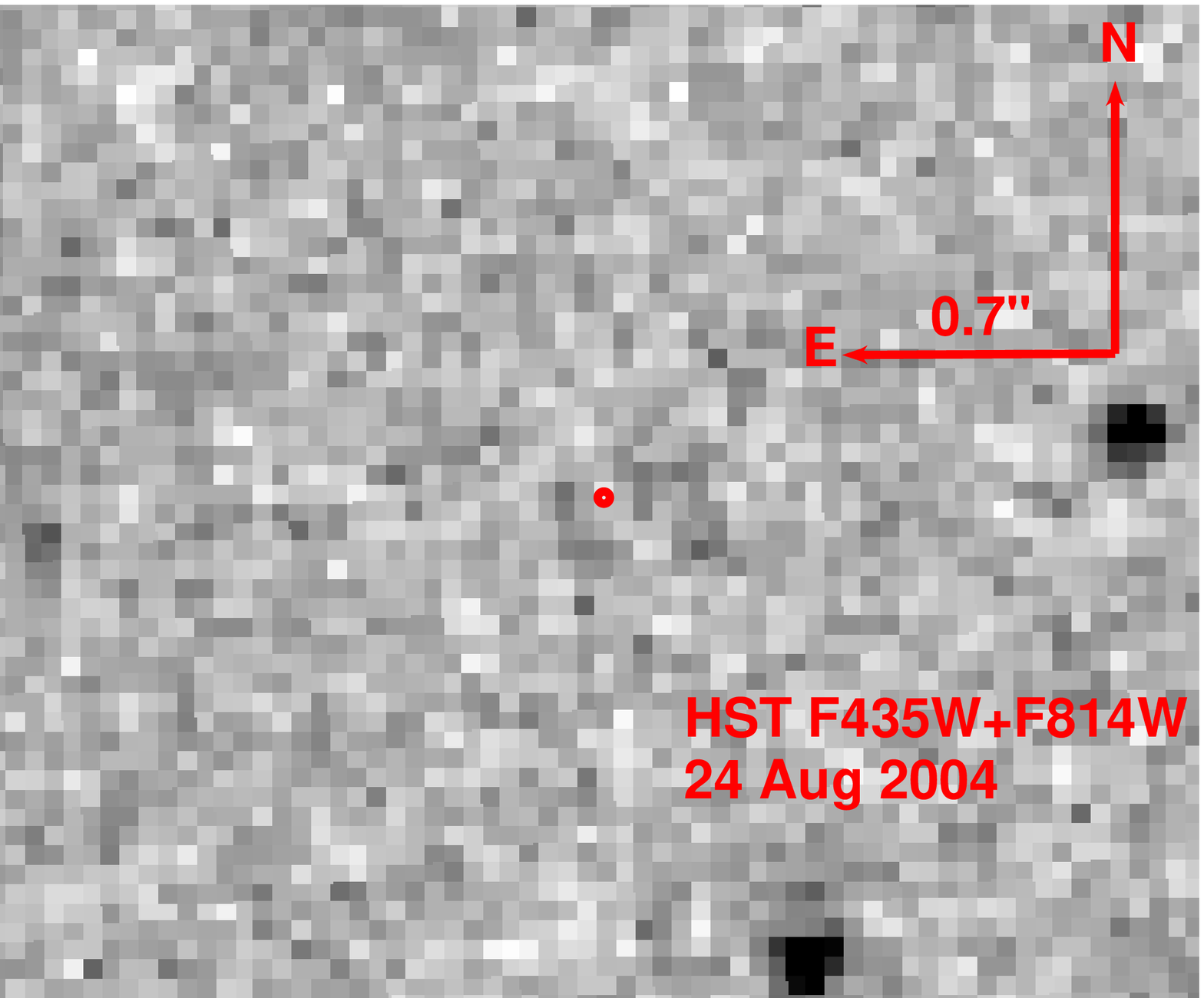}}
      \end{minipage}
      \begin{minipage}{6.8in}
            \fbox{\includegraphics[width=0.321\textwidth]{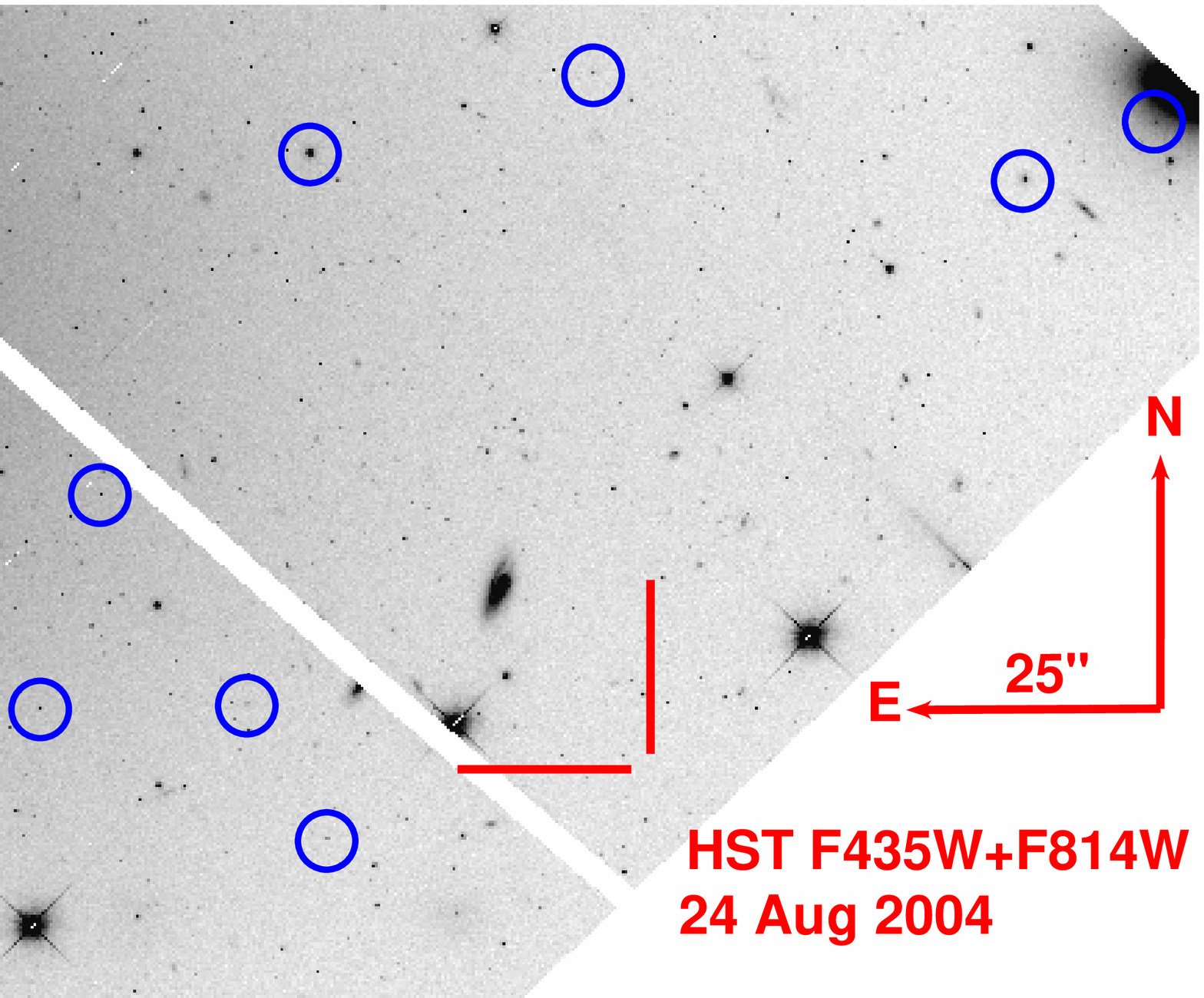}}
      \fbox{\includegraphics[width=0.321\textwidth]{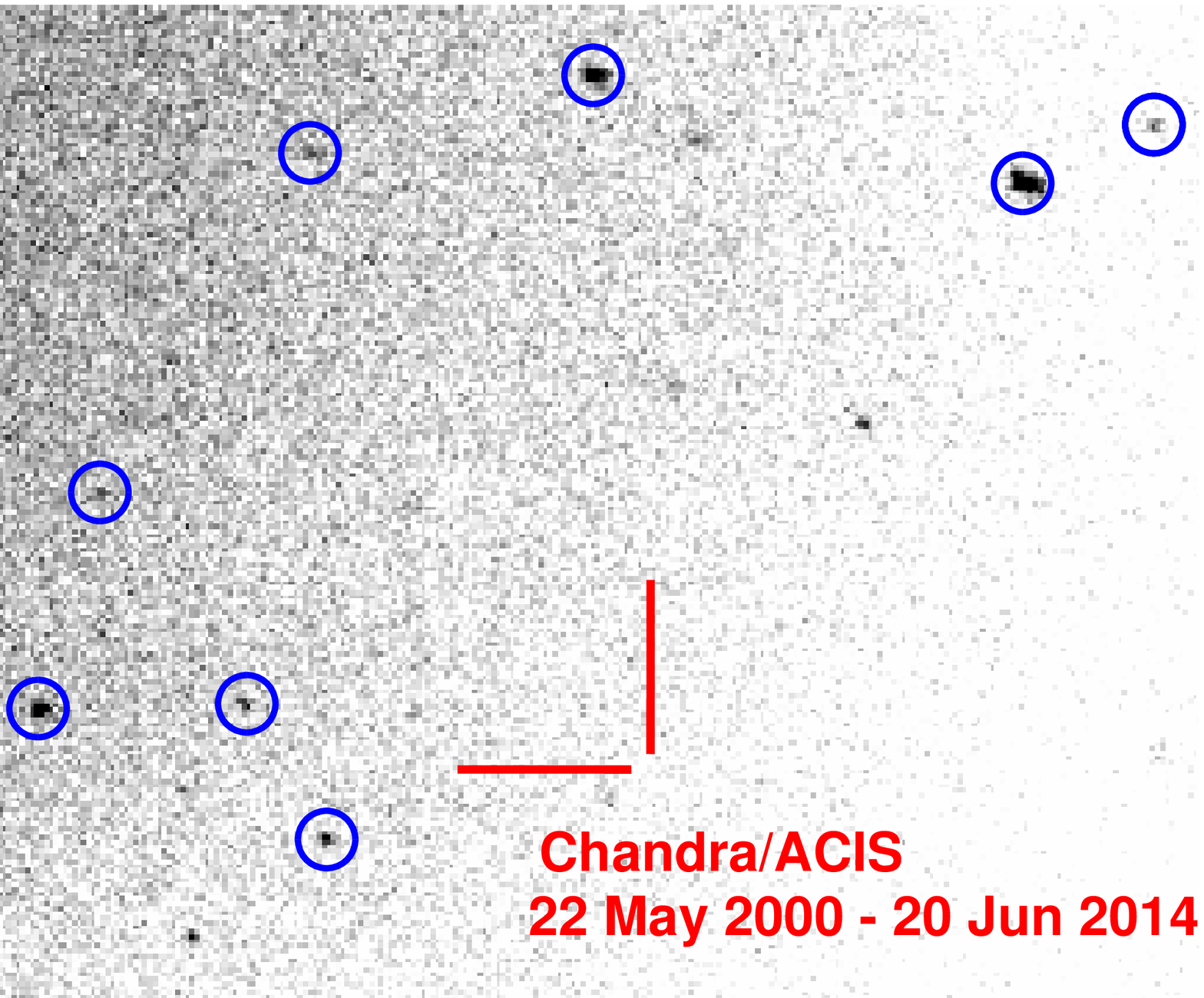}}
      \fbox{\includegraphics[width=0.321\textwidth]{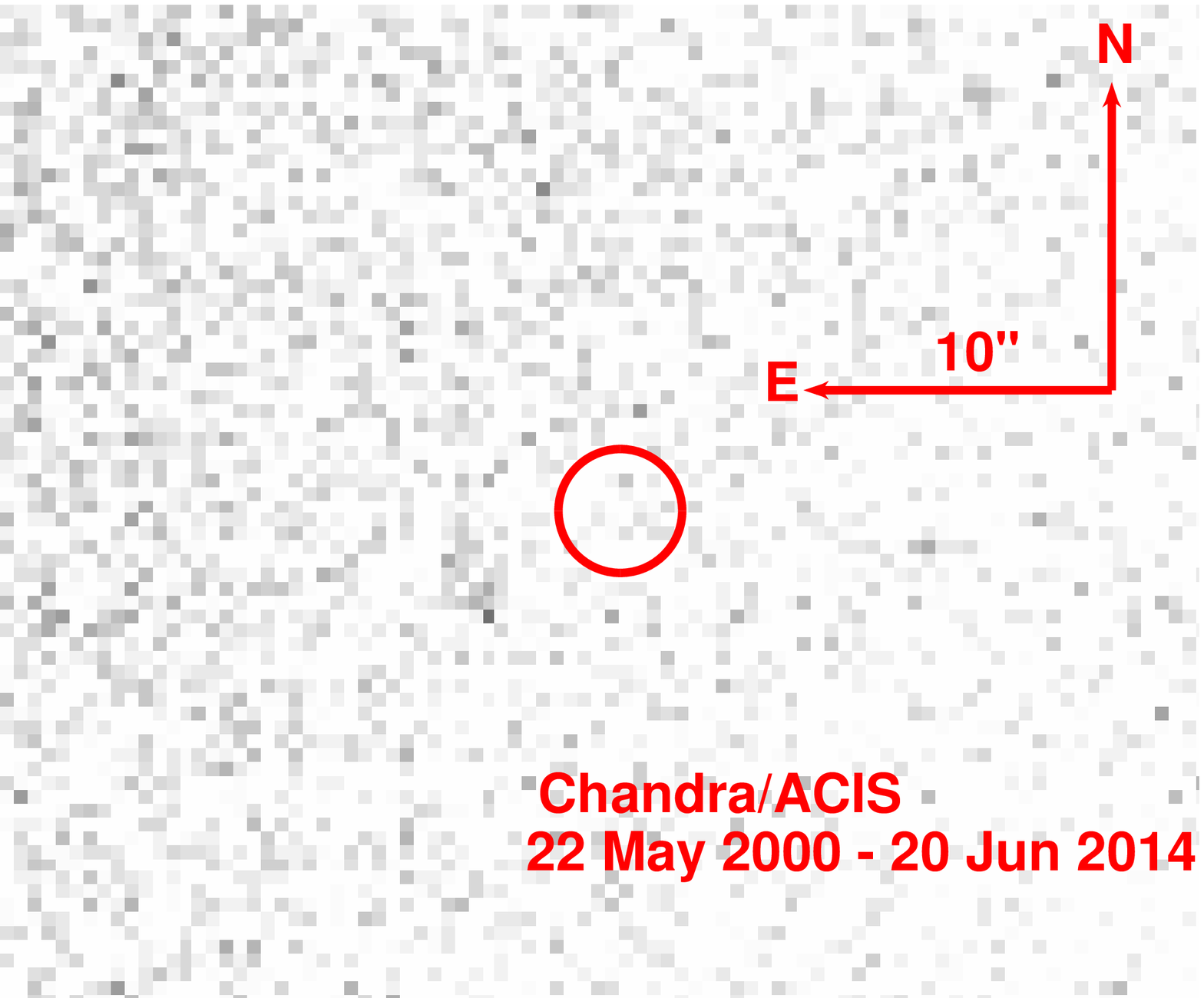}}
      \end{minipage}
      \end{center}
      \setlength{\fboxrule}{0.5pt}
      \caption{({\it Top Left}) Swope $B$-band image from 21 Jul. 2017 showing SN~2017ejb (red lines) relative to NGC~4696.  We circle $36$ sources used for relative astrometry with the \hst\ image in red.  ({\it Top Middle}) \hst/ACS $F435W$+$F814W$ image of NGC~4696 showing the same region as the image on the left.  The location of SN~2017ejb is denoted with red lines.  We circle the same $36$ sources in the Swope image.  ({\it Top Right}) A zoom-in of the \hst/$F435W$+$F814W$ panel in the middle.  We denote the location of SN~2017ejb as determined from relative astrometry with a red circle.  The size of the circle corresponds to our astrometric uncertainties ($\approx0.016\arcsec$).  There are no sources in the \hst\ image within $>72\sigma$ of the location of SN~2017ejb.  ({\it Bottom Left}) The same \hst/$F435W$+$F814W$ image as above.  We circle $8$ sources in blue used for relative astrometry.  We mark the location of SN~2017ejb with red lines.  ({\it Bottom Middle}) {\it Chandra}/ACIS image of the same region on the left.  We circle the same $8$ sources in the \hst\ image in blue and mark the location of SN~2017ejb.  ({\it Bottom Right}) A zoom-in of the {\it Chandra}/ACIS image in the middle showing a $4.5$~pixel region centered on the location of SN~2017ejb.  We do not detect any point-like sources at the $>3\sigma$ level in this region.}\label{fig:images}
\end{figure*}

\subsubsection{{\it Hubble Space Telescope}}\label{s:hst}

The site of SN~2017ejb was also observed by the {\it Hubble Space Telescope} (\hst) with the Advanced Camera for Surveys (ACS) Wide Field Channel (WFC) in $F435W$ and $F814W$.  These images were observed over a single epoch on 24 Aug. 2004.  We obtained the individual {\tt flc} files from the the Mikulski Archive for Space Telescopes\footnote{\url{https://archive.stsci.edu/}}. These consisted of $4\times1360$~s exposures in $F435W$ and $4\times580$~s exposures in $F814W$.  Following procedures described in \citet{kilpatrick+18}, we drizzled the images together and performed photometry on the {\tt flc} files using {\tt dolphot} \citep{dolphin00}.  We used standard {\tt dolphot} parameters for ACS\footnote{\url{http://americano.dolphinsim.com/dolphot/dolphotACS.pdf}}.  The instrumental magnitudes were calibrated using the zero points for \hst/ACS from 24 Aug. 2004\footnote{\url{https://acszeropoints.stsci.edu/}}.  For reference to the individual {\tt flc} files, we drizzled all $F435W$ and $F814W$ together to construct the deepest image possible ($F435W$+$F814W$), which is shown in \autoref{fig:images}.

\subsection{Spectroscopy}\label{s:spectroscopy}

\begin{figure}
      \includegraphics[width=0.47\textwidth]{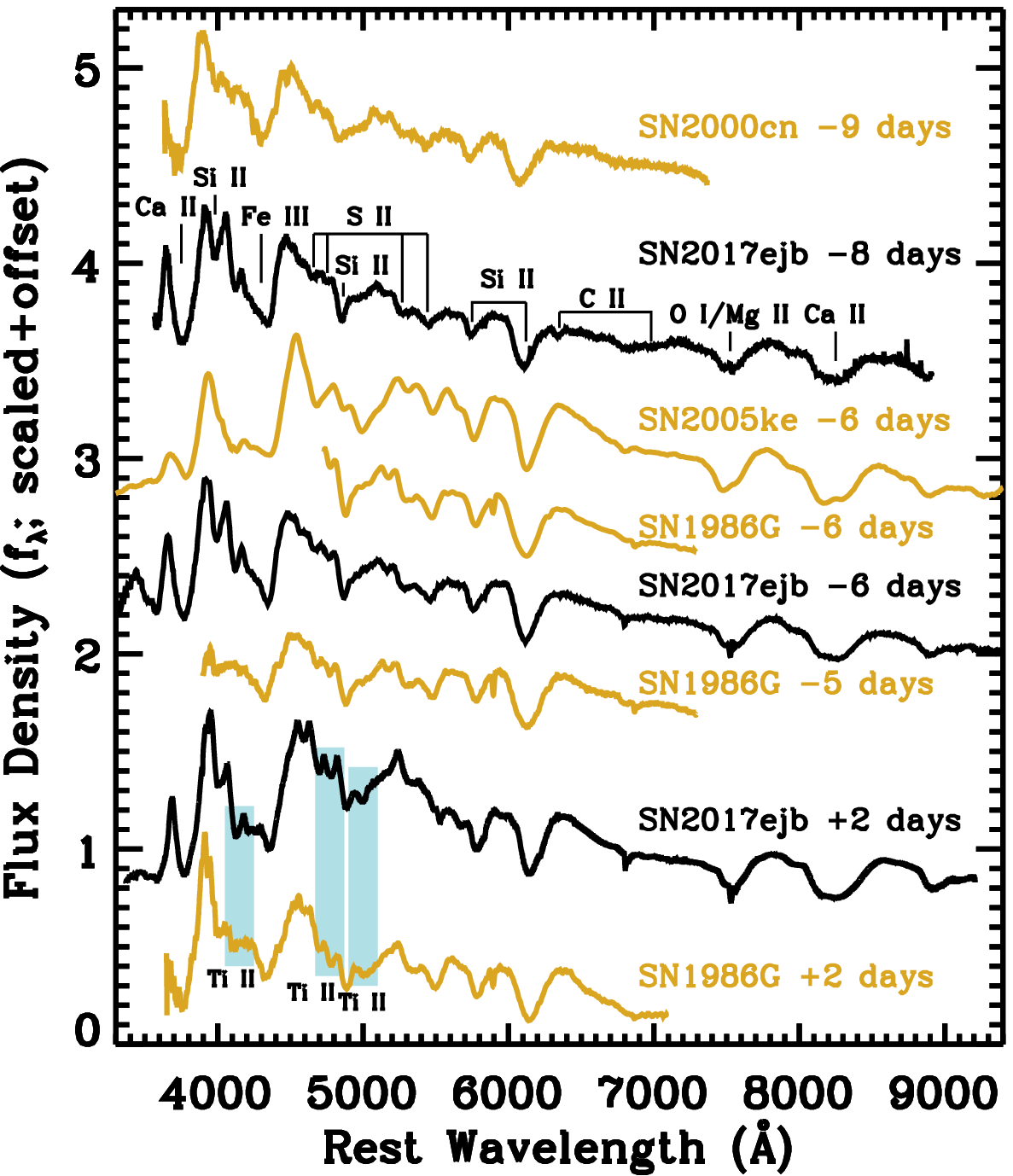}
      \caption{Spectra of SN~2017ejb (black) labeled with the time of observation relative to $B$-band maximum.  In our first epoch, we label several spectroscopic features present.  For comparison, we overplot spectra of the low-luminosity SN~Ia 2000cn \citep{matheson+08}, the 1991bg-like SN~Ia 2005ke \citep{folatelli+13}, and the low-luminosity SN~Ia 1986G \citep{phillips+87}.  All spectra have been de-reddened for Milky Way and host reddening and their recessional velocities have been removed.  We note the presence of Ti\II\ bands (shaded blue) in the SN~2005ke spectrum from $6$~days before $B$ maximum, which are either weak or missing in the SN~2017ejb spectra.}\label{fig:spectra}
\end{figure}

We observed SN~2017ejb on 29.04 May 2017 with the Goodman Spectrograph \citep{clemens+04} on the 4.1~m Southern Astrophysical Research Telescope (SOAR) on Cerro Pach\'{o}n, Chile. Our SOAR/Goodman setup and spectral reduction procedure are described in \citet{kilpatrick+18}.  We de-reddened the spectrum for the Milky Way value and removed the recession velocity $2960$~km~s$^{-1}$, which is consistent with the redshift of NGC~4696. This spectrum is shown in \autoref{fig:spectra}.

SN~2017ejb was also observed on 1.18 Jun 2017 with the ESO Faint Object Spectrograph and Camera (EFOSC2) on the ESO 3.6~m New Technology Telescope (NTT) at La Silla Observatory, Chile as part of the PESSTO programme\footnote{\url{www.pessto.org}} \citep[for a description of the observing programme and instrumental setup, see][]{smartt15}.  We reduced these data following standard procedures in {\tt IRAF}\footnote{IRAF, the Image Reduction and Analysis Facility, is distributed by the National Optical Astronomy Observatory, which is operated by the Association of Universities for Research in Astronomy (AURA) under cooperative agreement with the National Science Foundation (NSF).}.  The final spectrum is shown in \autoref{fig:spectra}.

We also obtained a spectrum observed with X-shooter on the Very Large Telescope (VLT) on Cerro Paranal, Chile on 9.03 June 2017\footnote{from \url{http://archive.eso.org/cms.html}} (ESO programme 099.D-0641, PI Maguire).  The data were processed using the latest version of the X-shooter pipeline \citep{modigliani+10} with calibration frames and standard star spectra obtained on the same night and in the same instrumental configuration.  We combined data from the ultraviolet/blue, optical, and infrared arms of X-shooter by scaling the individual spectra to the overlap region between each side.  We show the combined spectrum in \autoref{fig:spectra}.

\subsection{Swope Imaging}\label{s:swope}

We observed SN~2017ejb using the Direct CCD Camera on the Swope 1.0~m telescope at Las Campanas Observatory, Chile, between 4 Jun. 2017 and 16 Aug. 2017 in \ubvgri\footnote{Swope filter functions are provided at \url{http://csp.obs.carnegiescience.edu/data/filters}}. We performed standard reductions on the Swope data, including bias-subtraction, flat-fielding, cross-talk correction, astrometry, and photometry, using the {\tt photpipe} imaging and photometry package \citep{rest+05} as discussed in \citet{kilpatrick+18}.  We did not subtract a template from images with the SN, but we accounted for the sky and host galaxy background level by fitting to the median background level around the PSF aperture.

\begin{figure}
      \includegraphics[width=0.47\textwidth]{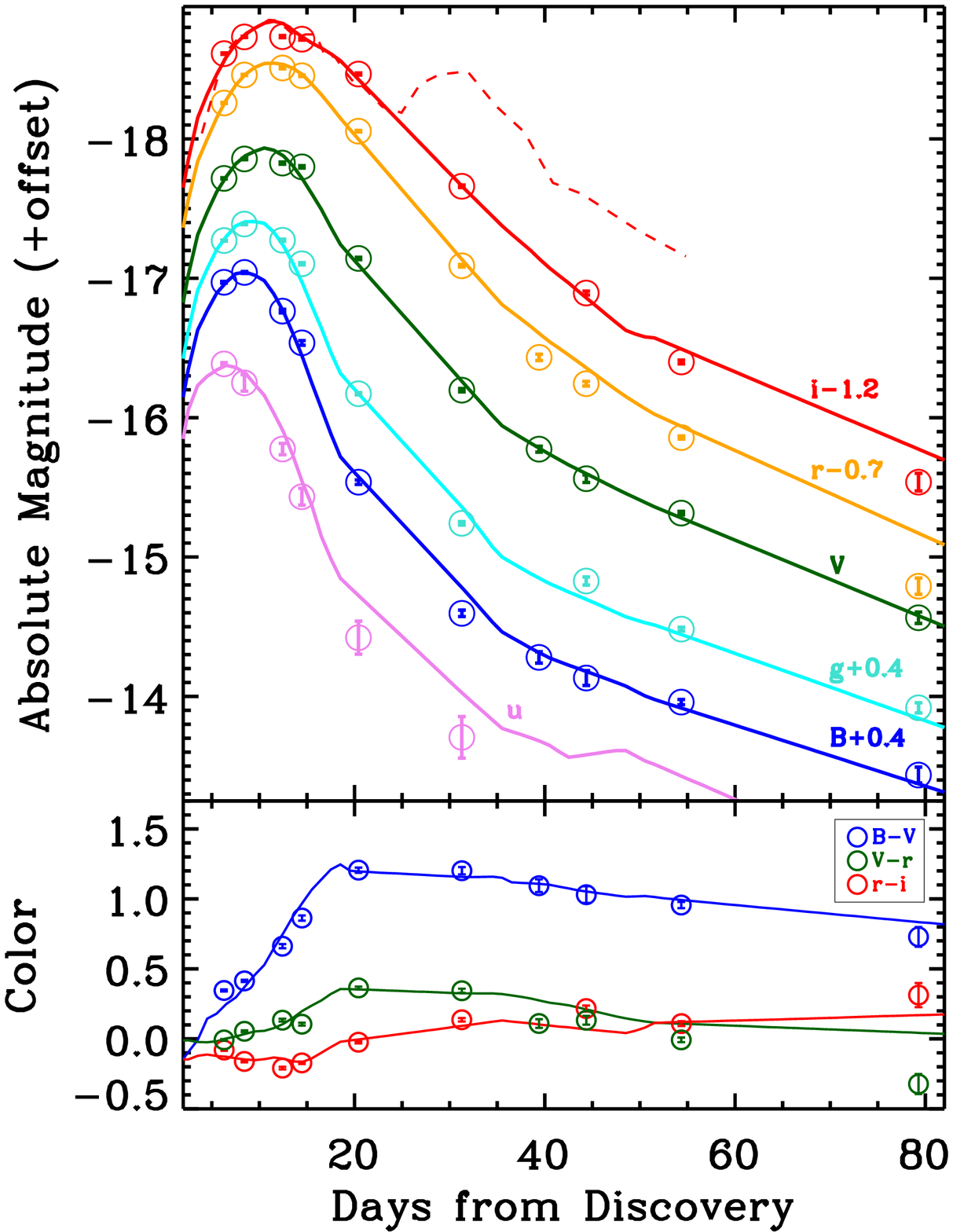}
      \caption{({\it Top}) Swope $uBV\!gri$ light curves of SN~2017ejb (circles), which have been corrected for Milky Way reddening and shifted to the distance of NGC~4696.  For comparison, we overplot Swope in-band SiFTO light curves \citep[solid lines;][]{conley+08} based on a SN~1991bg template \citep{nugent+02} and fit to the data of SN~2017ejb.  We also plot the $I$-band light curve (dashed red line) of the normal but low-luminosity SN~Ia 2007au \citep{ganeshalingam+10}.  All comparison light curves have been corrected for Milky Way extinction and shifted to the distances in their respective references.  ({\it Bottom}) $B-V$, $V-r$, and $r-i$ color curves of SN~2017ejb (red, green, and blue circles, respectively).  We also overplot the SiFTO color curves from the fits in the top panel.}\label{fig:lc}
\end{figure}

We calibrated the $ugri$ photometry using SkyMapper secondary standards \citep{wolf+18} in the same field as SN~2017ejb.  For our $BV$ photometry, we transformed the SkyMapper standard star $gr$ magnitudes to $BV$ using transformations in \citet{jester+05}.  SN~2017ejb was clearly detected in each epoch at the coordinates reported in \citet{atel10439}.  The final photometry of SN~2017ejb is presented in \autoref{tab:phot} and shown in \autoref{fig:lc}. 

\begin{table*}
\begin{center}\begin{minipage}{4.2in}
      \caption{Swope Optical Photometry of SN~2017ejb}\small
      \setlength\tabcolsep{2.5pt}
\begin{tabular}{@{}lcccccc}\hline\hline
Epoch & $u$ & $B$ & $V$ & $g$ & $r$ & $i$ \\ \hline
6.82  & 16.844 (008) & 15.785 (006) & 15.439 (004) & 15.404 (003) & 15.447 (003) & 15.527 (004) \\
8.96  & 16.980 (059) & 15.714 (006) & 15.300 (004) & 15.286 (004) & 15.246 (004) & 15.407 (004) \\
12.96 & 17.455 (040) & 15.992 (012) & 15.329 (007) & 15.402 (007) & 15.196 (006) & 15.405 (007) \\
15.00 & 17.797 (060) & 16.219 (016) & 15.356 (008) & 15.571 (008) & 15.252 (005) & 15.423 (005) \\
20.95 & 18.809 (118) & 17.219 (016) & 16.014 (008) & 16.503 (007) & 15.649 (004) & 15.673 (006) \\
31.81 & 19.524 (148) & 18.160 (023) & 16.959 (012) & 17.433 (012) & 16.615 (008) & 16.479 (008) \\
39.93 & ---          & 18.476 (040) & 17.382 (021) & 17.836 (039) & 17.272 (022) & 17.090 (016) \\
44.89 & ---          & 18.624 (053) & 17.594 (027) & 17.847 (029) & 17.461 (016) & 17.245 (013) \\
54.88 & ---          & 18.797 (018) & 17.841 (012) & 18.192 (013) & 17.848 (010) & 17.739 (013) \\
79.81 & ---          & 19.320 (057) & 18.592 (040) & 18.757 (036) & 18.914 (058) & 18.602 (063) \\
\hline
\end{tabular}\label{tab:phot}
\begin{tablenotes}
      \small
\item {\bf Note.} Epoch is in days relative to discovery on 28.22 May 2017. Uncertainties ($1\sigma$) are in millimagnitudes and given in parentheses next to each measurement. All photometry is on the AB scale.
\end{tablenotes}
\end{minipage}
\end{center}
\end{table*}

\section{Photometric and Spectral Classification of SN~2017ejb}\label{s:classification}

\subsection{Spectroscopic Classification}\label{sec:spectral}

In \autoref{fig:spectra}, we show all of our spectral epochs of SN~2017ejb with several spectroscopic features identified.  At $8$~days before $B$-band maximum (as determined in \autoref{sec:lc}), our SN~2017ejb spectrum exhibits prominent lines of Si\II, S\II, Ca\II, and C\II, which indicate that SN~2017ejb is a SN~Ia.  We only detect C\II\ absorption in our first spectroscopic epoch, roughly $8$~days before $B$-band maximum.  While the presence of the C\II\ $\lambda$6580 feature and possible detection of C\II\ $\lambda$7234 in SN~2017ejb is not unprecedented \citep[see, e.g., full analysis of C\II\ features in SNe~Ia in][]{parrent+11}, SNe with spectra $>1$~week before maximum light and strong C\II\ absorption are rare.  C\II\ absorption is often an indicator that the SN~Ia has other peculiarities, such as in the extremely low-luminosity SN~Ia 2008ha \citep{foley+09} or the low-velocity SN~Ia 2009dc \citep{taubenberger+11a}.

In our first spectroscopic epoch, the C\II\ features are blueshifted to the same velocity of $-$11,200$\pm300~\text{km s}^{-1}$, which is comparable to the Si\II\ $\lambda$6355 velocity of $-$11,900$\pm200~\text{km s}^{-1}$ (i.e., with a C\II\ to Si\II\ velocity ratio of $0.94\pm0.04$).  This ratio is low, although nominally consistent with the population of SNe~Ia studied in \citet{parrent+11}, and comparable to specific examples such as SNe~1994D and 1996X \citep{patat+96,salvo+01}.

For comparison to our SN~2017ejb spectra in \autoref{fig:spectra}, we plot spectra of other peculiar or low-luminosity SNe~Ia at similar epochs with respect to $B$-band maximum, including SN~2000cn \citep{matheson+08}, SN~1986G \citep{phillips+87}, and SN~2005ke \citep{matheson+08}.  All of the comparison spectra have been de-reddened and the recessional velocity of the SN host galaxy has been removed according to the values in each reference.  SN~2017ejb shares similarities with all of these objects, especially the velocity and relative ratio of Si\II\ features.  

In the VLT/X-shooter spectrum at $+$2~days after $B$-band maximum, SN~2017ejb exhibits strong, broad Ti\II\ bands characteristic of 1986G-like SNe near $\lambda$4650 and 5000 \citep[][also see labels in \autoref{fig:spectra}]{phillips+87}.  These lines are much more prominent near peak light than in our pre-maximum spectra.  This finding is consistent with the presence of Ti\II\ in SN~Ia spectra overall, which is an indication of relatively low ejecta temperatures \citep{doull+11} where SNe with hotter ejecta have Ti in higher ionization states with fewer and weaker absorption features in the optical.

Finally, although we detect Na\I\ D absorption from the Milky Way with an equivalent width (EW) of $0.7\pm0.1$~\AA\ \citep[which is consistent with the Milky Way reddening of $E(B-V)=0.098$~mag using the relation in][]{poznanski+12}, we do not detect any Na\I\ D extinction at the redshift of NGC~4696 at the $<0.1$~\AA\ level in any of our spectra.  This implies a host reddening of $E(B-V)<0.02$~mag.  NGC~4696 does have an extended, relatively massive dust lane that was likely captured $10^{8}$~yr ago \citep{dejong+90,sparks+89}. However, SN~2017ejb is $158\arcsec$ (28~kpc at the distance of NGC~4696) in projection from the center of its host galaxy.  It is unlikely that it would be enshrouded by significant host extinction.

\subsection{Light Curves}\label{sec:lc}

Assuming the distance and Milky Way extinction above, we plot the extinction-corrected absolute magnitudes for SN~2017ejb in \autoref{fig:lc}.  The key characteristics of SN~2017ejb are its low peak magnitude, rapid decline, and apparent lack of a secondary $i$-band maximum.  The $B$-band light curve peaks around Julian Date 2457910.8$\pm$1.0 with $M_{B}=-17.9\pm0.1$~mag and declines with $\Delta m_{B,15}=1.7\pm0.1$~mag.  This $\Delta m_{B,15}$ value corresponds to a light curve stretch parameter $x_{1}\approx-2.9$ \citep[see, e.g.,][]{guy+07}, which is comparable to many low-luminosity SNe~Ia \citep[see distributions in, e.g.,][]{hicken+09}.

Only a small minority of normal SNe~Ia have light curve parameters $x_{1} < -2.9$ or $\Delta m_{B,15} > 1.7$~mag \citep[e.g., only SNe~1998co and 2007cp out of 146 SNe~Ia used for cosmology in][]{rest+14}.  This is partly by design as light curve fitting schemes for cosmology only yield accurate distances for $|x_{1}|<3.0$ \citep[e.g., SALT;][]{guy+10,betoule+14}, and SNe outside this range are typically cut from cosmological samples \citep[see, e.g., the homogeneous low-redshift sample in][]{foley+18}.  There are many low-luminosity SNe~Ia such as SNe~1986G, 1991bg, 1993H, and 1999by that peak around $M_{B}=-$16.5 to $-$18.0~mag and exhibit $\Delta m_{B,15}=1.7$--$2.0$~mag \citep{phillips+87,filippenko+92,garnavich+01,altavilla+04}. In this regard, SN~2017ejb appears to fall between the distribution of normal SNe~Ia used for cosmology and low-luminosity SNe~Ia.

One unusual feature in the SN~2017ejb light curve compared with normal SNe~Ia is the apparent lack of a second peak in the $i$-band light curve, which SNe~Ia typically exhibit $20$--$30$~days after peak \citep[even for normal SNe~Ia with low values of $x_{1}$;][]{kasen+06}.  For comparison, we plot the $I$-band light curve of the low-luminosity but normal SN~Ia 2007au \citep{ganeshalingam+10}, which does exhibit a secondary $I$-band maximum (dashed red line in \autoref{fig:lc}).  SN~2007au has $M_{B,{\rm peak}}=-18.0$~mag and very similar light curve parameters to SN~2017ejb \citep[SN~2007au has $x_{1}=-2.82$ from][]{rest+14}.  Thus, if SN~2017ejb had been a photometrically normal SN~Ia, it is reasonable to expect that it would have had a prominent secondary $i$-band maximum that would be apparent in our data. SN~2017ejb is more similar to peculiar, low-luminosity SNe~Ia in this regard, such as SNe~1986G and 1991bg \citep{phillips+87,filippenko+92}, although with a $\Delta m_{B,15}$ parameter that is on the low end for this population.

In order to find the best-matching light curve template for SN~2017ejb, we performed SiFTO light curve fits \citep{conley+08} using a 1991bg-like template from \citet{nugent+02}.  We used the Swope filter functions to generate in-band light curves matched to the observed data from SN~2017ejb (\autoref{fig:lc}).  The fits are relatively good before and around maximum light, but diverge $50$~days into the post-maximum phase (in $r$- and $i$-bands) and in $u$-band generally where most SN~Ia light curves are poorly constrained \citep[especially low-luminosity SNe~Ia;][]{taubenberger+08}.

This overall similarity with a peculiar sub-class of SNe~Ia, the low peak luminosity, the rapid decline rate compared with most normal SNe~Ia, and the lack of a secondary $i$-band maximum seem to indicate that SN~2017ejb is a member of the peculiar, low-luminosity class of SNe~Ia such as SNe~1986G and 1991bg.  These evidence reinforce the spectroscopic similarity between SN~2017ejb and SN~1986G.

\section{Pre-explosion Limits on a Counterpart to SN~2017ejb}\label{s:limits}

\subsection{Relative Astrometry and \hst\ Limits}

We examined the {\it Chandra}/ACIS data described above near the explosion site of SN~2017ejb.  In order to place constraints on the total number of events associated with the SN~2017ejb progenitor system in the {\it Chandra} data, we must precisely constrain the location of the explosion site in the {\it Chandra} images.  Archival {\it Chandra} data products are astrometrically calibrated using the Tycho-2 \citep{hog+00}, USNO-A2.0 \citep{urban+98}, and 2MASS \citep{skrutskie+06} astrometric catalogs. Similarly, we reduce Swope optical imaging in {\tt photpipe} using 2MASS astrometric standards as described in \citet{kilpatrick+18}. However, we cannot rule out the possibility that there is some systematic offset between {\it Chandra} and Swope astrometry, and so we cross-checked our astrometric calibration by performing relative astrometry between the {\it Chandra} and Swope imaging using field sources.

This process is complicated by the fact that bright, compact X-ray sources tend to be extremely faint or extended in optical imaging. Therefore, we aligned our $B$-band image of SN~2017ejb at peak (the $8.96$~day epoch in \autoref{tab:phot}) to the drizzled $F435W$+$F814W$ \hst\ image and bootstrapped the relative astrometry to the {\it Chandra} event map.  This process is relatively straightforward given that our Swope $B$-band image covers roughly the same wavelengths as \hst/ACS $F435W$.

Identifying $36$~sources common to our stacked Swope $B$-band image and drizzled \hst\ $F435W$+$F814W$ image, we performed relative astrometry between the two images.  We estimated the uncertainty in our astrometric solution by randomly selecting $18$ of these sources and calculating an astrometric solution, then calculating the average offset between the remaining $18$ sources.  Repeating this process, we estimated the average offset between these sources to be $\sigma_{\alpha}=0.014\arcsec$ and $\sigma_{\delta}=0.013\arcsec$.  We then determine the location of SN~2017ejb using coordinates from {\tt photpipe}.  SN~2017ejb is detected at $\sim$160$\sigma$ in the Swope $B$-band image with a full-width at half-maximum of $1.3\arcsec$, and so we estimate that the approximate location of the SN contributes $\approx$0.008\arcsec\ to the astrometric uncertainty. At the location of SN~2017ejb, we do not detect any sources at the $\geq3\sigma$ level in the individual or stacked \hst\ images.  The closest source of any kind is detected at $8.9\sigma$ in the drizzled $F435W$+$F814W$ image and is 1.15\arcsec\ away from the location of SN~2017ejb, or about 72 times the total astrometric uncertainty (\autoref{fig:images}).  Thus, we conclude that there is no source in any of these images consistent with being the progenitor system of SN~2017ejb.

We then identified $8$ sources common to the drizzled \hst\ image and {\it Chandra} image (\autoref{fig:images}).  Using {\tt sextractor} to determine the centroids of these sources in the {\it Chandra} image, we repeated the same process above, with $4$ sources to calculate a WCS solution and using the remaining $4$ sources to estimate the average offset.  We found an average offset of $\sigma_{\alpha}=0.10\arcsec$ and $\sigma_{\delta}=0.08\arcsec$.  Therefore, the combined uncertainty in the position of SN~2017ejb in the {\it Chandra} image using our Swope$\rightarrow$\hst$\rightarrow${\it Chandra} relative astrometry is approximately $0.11\arcsec$, or roughly 0.22 {\it Chandra} pixels.

We also used {\tt dolphot} to estimate the $3\sigma$ limiting magnitude on the presence of a source in the $F435W$ and $F814W$ images.  Using the {\tt FakeStar} parameter, we injected 10000 sources with a fixed number of counts into the {\tt flc} files.  We chose the positions for each of these sources by generating Gaussian random variables $x,y$ centered at the best-fitting pixel coordinates of SN~2017ejb and with standard deviations corresponding to the astrometric uncertainty in our relative astrometry on the location of SN~2017ejb ($0.32$ ACS/WFC pixels).  We increased the number of counts associated with these sources and repeated the process until we recovered $\geq9970$ sources at the $\geq3\sigma$ level.  In this way, we determined that the $3\sigma$ limiting magnitude on the presence of a point source at the location of SN~2017ejb to be $m_{F435W} > 28.3$~mag and $m_{F814W} > 26.8$~mag.

For the distance and Milky Way extinction to NGC~4696, the \hst\ limits correspond to $M_{F435W} > -5.2$~mag and $M_{F814W} > -6.4$~mag.  Even for a relatively small bolometric correction (e.g., $BC_{F435W} = 0$), the $F435W$ limiting magnitude corresponds to a source with $\log(L/L_{\odot})=4.0$, which is approximately the luminosity of a $13~M_{\odot}$ main-sequence star based on Mesa Isochrone \& Stellar Track evolutionary models \citep{paxton+11,paxton+13,paxton+15,dotter+16,choi+16}\footnote{\url{http://waps.cfa.harvard.edu/MIST/}}.  For stars with redder colors (i.e., where the $F814W$ bolometric correction is small), we can rule out stars with $\log(L/L_{\odot})=4.5$, which corresponds $M_{init}=10$--$13~M_{\odot}$ red supergiants.  These limits are not very constraining in the context of SN~Ia progenitor systems --- high-mass stars in this range would explode before a WD could evolve.

\subsection{{\it Chandra} Limits}\label{s:chandra-lim}

Although there are events detected near the location of SN~2017ejb in the {\it Chandra} image, this emission is smooth and likely associated with the hot gas surrounding NGC~4696 \citep[as analyzed in, e.g.,][]{crawford+05,fabian+05}.  Following methods described in \citet{nielsen+12}, we considered the total number of counts within a 4.5~pixel radius of the location of SN~2017ejb as this is where $>$95\% of the energy is encircled for a point source observed by {\it Chandra}/ACIS\footnote{see \url{http://cxc.harvard.edu/proposer/POG/html/}}.  Within a 4.5~pixel radius of the location of SN~2017ejb, we detected a total of 509~counts in the 0.3--1.0~keV {\it Chandra}/ACIS soft bandpass (\autoref{fig:images}).  There is no evidence for a point-like source at this location.

As in \citet{gehrels+86} and \citet{nielsen+12}, we calculated the maximum average number of counts $\mu$ for which the probability of observing $x \leq N$ counts (where, here, $N=509$) is within $3\sigma$ (i.e., $P(\mu; x \leq N) \leq 0.0013$).  Since the observed number of counts $N$ is large, we approximated the value of $\mu$ using equation (9) in \citet{gehrels+86} for a $3\sigma$ limit to be $\mu\approx581$. This value represents the maximum $3\sigma$ limit on a $0.3$--$1.0$~keV source at the location of SN~2017ejb, which includes background counts.

In order to remove the contribution from background counts, we calculated the number of counts per pixel around the location of SN~2017ejb using an annulus with inner radius 9~pixels and outer radius 18~pixels.  The average number of counts per pixel is 8.02~ct~pixel$^{-1}$, and so we approximated the maximum number of counts for a $3\sigma$ detection of a source at the location of SN~2017ejb to be $\mu^{\prime}\approx 581 - \pi \times (4.5)^{2} \times 8.02 = 70.8$.  This value is roughly consistent with the $3\sigma$ limit derived by assuming that the source is entirely dominated by Poisson noise from the background (i.e., $3 \times \sqrt{\pi \times (4.5)^{2} \times 8.02} = 67.8$).  Therefore, we are confident that 70.8~ct is a conservative $3\sigma$ limit on the total 0.3--1.0~keV counts from any pre-explosion counterpart to SN~2017ejb.

The flux limit in the 0.3--1.0~keV band depends on the effective exposure map at the location of SN~2017ejb for the merged {\it Chandra}/ACIS data.  We generated a weighted exposure map by assuming that any source detected at the location of SN~2017ejb would have a 0.3--1.0~keV spectral profile resembling an absorbed blackbody.  Using \textsc{ciao}/{\tt xabsphot}, we modeled absorbed blackbodies with temperatures in the range $kT=20$--$200$~eV \citep[corresponding to the full range of observed SSS temperatures in, e.g.,][]{vandenheuvel+92,kahabka+97,ness+13}.

For the total column of hydrogen to NGC~4696, we note that the Milky Way extinction quoted above corresponds to $N_{H} = 6.76\times10^{20}~\text{cm}^{-2}$ using the best-fitting scaling relation in \citet{guver+09}. There is effectively zero host extinction to SN~2017ejb based on the absence of any Na\I\ D absorption in its optical spectrum, and so we do not account for any column of hydrogen in the host galaxy. However, we cannot rule out the possibility that there is circumstellar extinction originating from gas or dust around the progenitor system of SN~2017ejb and close enough that it would have been destroyed within the first few days after explosion \citep[i.e., before we obtained our spectrum such that the Na \I\ D is variable, as in][]{patat+07a,simon+09}.  We do not account for any such circumstellar extinction, but we acknowledge that this is a possibility for a WD accreting from a companion wind or in a symbiotic binary \citep[although it has been found that circumstellar extinction has little effect on the inferred X-ray luminosities for SSS temperatures $>30$~eV and accretion rates $<10^{-6}~M_{\odot}~\text{yr}^{-1}$;][]{nielsen+15}.

\begin{figure}
      \includegraphics[width=0.47\textwidth]{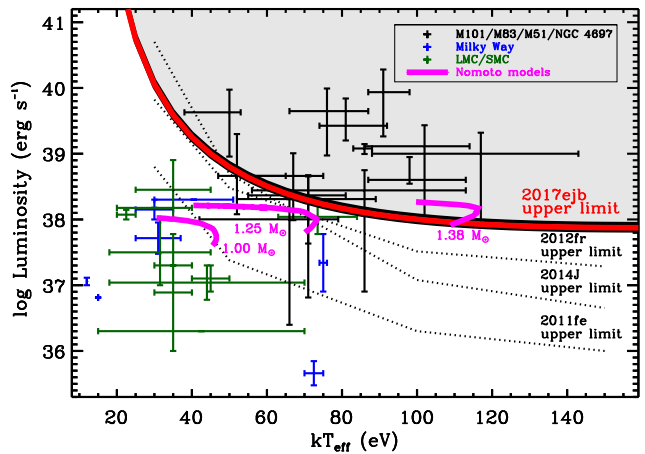}
      \caption{Hertzsprung-Russell diagram of supersoft X-ray sources (SSS).  We overplot the temperatures and luminosities of known SSS systems in M51, M81, M83, M101, NGC~4697 \citep[black; from][]{swartz+02,distefano+03}, the LMC, SMC, and Milky Way \citep[green and blue; from][]{greiner+00}.  For comparison, we overplot a model of a $1.38~M_{\odot}$ stably-accreting WDs (magenta lines) with masses $1.00$, $1.25$, and $1.38~M_{\odot}$ from \citet{nomoto+07}.  The derived limits on SSS systems of varying temperatures are shown for SNe~2011fe, 2012fr \citep{nielsen+13}, 2014J \citep{nielsen+14}, and 2017ejb (this paper).  We can rule out the hottest and most luminous SSS systems (grey region) as the progenitor of SN~2017ejb as well as the Chandrasekhar-mass WD accreting at rates $>3\times10^{-8}~M_{\odot}$.}\label{fig:ul}
\end{figure}

For every model spectrum, we calculated the value of the {\it Chandra}/ACIS exposure map at the location of SN~2017ejb ($\zeta$ in $\text{cm}^{2}~\text{s}$) and the average energy per photon in the 0.3--1.0~keV band ($\langle E \rangle$).  Thus, the $3\sigma$ upper limit on the 0.3--1.0~keV X-ray luminosity from the combined {\it Chandra} data is 

\begin{equation}
L_{X} = \frac{4 \pi \mu^{\prime} \langle E \rangle d^{2}}{\zeta}
\end{equation} 

\noindent for the values of the 3$\sigma$ count limit $\mu^{\prime}$ and distance $d$ given above.  In order to convert this upper limit to a bolometric luminosity, we calculated the fraction of the unabsorbed blackbody spectrum with temperature $T_{\rm eff}$ in the 0.3--1.0~keV band as 

\begin{equation} 
c(T_{\rm eff}) = \frac{\int_{0.3~\text{keV}}^{1.0~\text{keV}} B_{E}(T_{\rm eff}) dE}{\int_{0}^{\infty} B_{E}(T_{\rm eff}) dE}
\end{equation}

\noindent where $B_{E}(T)$ is the energy-dependent Planck function for a temperature $T$.  Thus, the upper limit on the bolometric luminosity for a model spectrum with effective temperature $T_{\rm eff}$ is $L_{\rm bol} = L_{X} / c(T_{\rm eff})$.  We show our upper limit (red) on the bolometric luminosity of any SSS counterpart to SN~2017ejb as a function of the assumed model temperature $kT_{\rm eff}$ in \autoref{fig:ul}.  We also show the effect of varying the distance to the Centaurus cluster within the $1\sigma$ uncertainties (black), which only has a marginal effect on the limiting luminosity.

\section{Discussion}\label{s:disc}

Our upper limit on the bolometric luminosity of any SSS is comparable to similar limits presented in \citet{nielsen+12} and \citet{nielsen+14} as shown in \autoref{fig:ul}.  In particular, the limits on a SSS counterpart at low temperatures are comparable to those for SNe~2014J and 2012fr, though not as constraining as for SN~2011fe.  For comparison, we show several known SSS systems \citep[from][]{greiner+00,swartz+02,distefano+03} as well as models for a Chandrasekhar-mass ($1.38~M_{\odot}$) and sub-Chandrasekhar mass ($1.00$ and $1.25~M_{\odot}$) stably-accreting SSS with acrretion rates $10^{-8}$ to $2.5\times10^{-7}~M_{\odot}~\text{yr}^{-1}$ from \citet{nomoto+07}.

Our $3\sigma$ limits for SN~2017ejb are $L_{\rm bol}=1.78\times10^{39}~\text{erg s}^{-1}$ at 40~eV, $L_{\rm bol}=3.20\times10^{39}~\text{erg s}^{-1}$ at 60~eV, and $L_{\rm bol}=1.00\times10^{38}~\text{erg s}^{-1}$ at 100~eV.  These limits rule out all known sources hotter than $kT_{\rm eff}=85$~eV and more luminous than $L_{\rm bol} = 4\times10^{38}~\text{erg s}^{-1}$.  The comparison SSS systems include a number of sources in M51, M81, M83, and M101 identified by \citet{swartz+02} and \citet{distefano+03} using {\it Chandra}, and so are systematically hotter and more luminous than sources identified, for example, in the Milky Way, LMC, and SMC using {\it ROSAT} \citep[][]{greiner+00}. We also rule out stably-accreting Chandrasekhar-mass WDs \citep[in particular, those in the temperature range expected for SSS systems;][]{nomoto+07} with the highest mass-loss rates ($>3\times10^{-8}~M_{\odot}~\text{yr}^{-1}$).

However, we cannot definitively rule out certain types of accreting WDs that lead to anomalously cool or low-luminosity SSS systems, for example, due to WD spin down \citep[e.g.,][]{distefano+11}.  In this scenario, an accreting WD can reach the Chandrasekhar mass but must spin-down and cool before it can explode.  In general, these scenarios are disfavored, as they would imply that most galaxies host a large population of rapidly-spinning WDs that will soon explode as SNe~Ia, which is not observed \citep[e.g.,][]{norton+04,ferrario+05}.

Another important caveat is that our most recent epoch of pre-explosion X-ray data was obtained roughly $3$~years before SN~2017ejb was discovered.  We are completely insensitive to any pre-explosion X-ray emission within those $3$~years. Furthermore, the bulk of the X-ray data ($486.25$~ks) were obtained around $3$~year before discovery, but we are significantly less sensitive to X-ray sources in this period than over the full $14$~years of observations.  If the SN~Ia ignition mechanism involves rapid mass transfer onto a WD on timescales comparable to or less than $\sim$3~years before explosion, we would not have detected any signature from that event.

We are also insensitive to circumstellar material in the immediate environment of the progenitor system that would be promptly destroyed and undetectable in the early-time spectra.  In general, such a large mass of material is not uexpected as optical and radio observations rule out large column densities of hydrogen in the immediate environments of SN~Ia progenitor systems \citep{leonard07,shappee+13,chomiuk+12,chomiuk+16}. Moreover, if a companion star to the WD progenitor of SN~2017ejb was losing mass at rates of $>10^{-6}~M_{\odot}~\text{yr}^{-1}$ \citep[i.e., where circumstellar extinction would have a significant effect on a SSS spectral profile;][]{nielsen+15}, then the WD would likely be accreting at a high rate and produce a luminous X-ray source.  We rule out stably accreting WDs with mass accretion rates around $3\times10^{-8}$--$2.5\time10^{-7}~M_{\odot}~\text{yr}^{-1}$, and so it is unlikely SN~2017ejb exploded from a system with an even higher accretion rates (and intrinsic luminosities) $>4$ times as large but some circumstellar extinction.

Overall, there is significant parameter space within which an accreting WD progenitor system to SN~2017ejb could have undergone a Chandrasekhar or sub-Chandrasekhar mass explosion.  If the progenitor WD had a sub-Chandrasekhar mass ($<1.35~M_{\odot}$), or had a low accretion rate ($<3\times10^{-8}~M_{\odot}~\text{yr}^{-1}$), or underwent rapid mass-transfer within the last few years before explosion, we would not have detected an X-ray source.  Any of these scenarios is plausible, but together they add context to the characteristics and large-scale environment of the SN~2017ejb explosion.

In particular, we note that SN~2017ejb was discovered in the massive elliptical galaxy NGC~4696 \citep{shobbrook+63,mitchell+75}.  While this galaxy exhibits tendrils of dust that likely originate from material captured $10^{8}$~yr ago, most of the star formation in NGC~4696 is suppressed by its central black hole \citep{sanders+16}.  SN~2017ejb is also at least 20~kpc in projection from these dust lanes, implying that if the progenitor system originated from a burst of star formation in this material, it must have had a projected velocity $\geq200~\text{km s}^{-1}$ very soon after it formed.  This scenario is plausible if the stars produced in this burst maintained some of the velocity from the infalling material.  On the other hand, the progenitor system could also have originated from a previous burst of star formation in NGC~4696 and before its central black hole became highly active.  This scenario would support the conclusions of studies such as \citet{howell+01} and \citet{piro+14}, who point to older stellar populations in galaxies with low star formation as likely sites for binary WD mergers. In addition, our findings support the hypothesis that some low-luminosity SNe~Ia may be the result of binary WD mergers \citep[][]{pakmor+09}.

Combined with the growing sample of optical and X-ray limits in the literature \citep{nelemans+08,maoz+08,li+11,nielsen+13,nielsen+14,kelly+14}, the SN~2017ejb limits rule out interesting regions in WD temperature and luminosity for plausible progenitor systems.  In particular, we can rule out most systems near the Chandrasekhar limit, assuming they were stably accreting for years before explosion.  Combined with the limits from SNe~2011fe, 2012fr, and 2014J, we conclude at the 95\% confidence level that $<$47\% of SNe~Ia explode from systems involving a stably-accreting Chandrasekhar-mass SSS.  

Future analysis of pre-explosion imaging for all SNe~Ia can be used to verify or constrain expectations for the configuration of their progenitor systems and explosion scenarios.  In light of the wide variety of SN~Ia explosion models, this type of analysis provides one of the most promising lines of inquiry for resolving the SN~Ia progenitor problem.

\section{Conclusions}\label{s:conc}

We analyze post-explosion imaging and spectroscopy of SN~2017ejb and pre-explosion \hst\ and {\it Chandra} imaging of its explosion site.  In summary, we find:

\begin{enumerate}
      \item SN~2017ejb is a low-luminosity SN~Ia with strong C\II\ absorption features in its pre-maximum spectra.  Photometrically, it has a low peak luminosity, it declines quickly, and it lacks a secondary $i$-band maximum.  Spectroscopically, it is similar to SN~1986G, but with relatively weak Ti\II\ bands.  Overall, it is most similar to low-luminosity SNe~Ia such as SNe~1986G and 1991bg.
      \item We do not detect any counterpart to SN~2017ejb in pre-explosion {\it Chandra} imaging.  Assuming that any pre-explosion {\it Chandra} source resembles a blackbody obscured by Milky Way extinction, our limits correspond to $L_{\rm bol} = 4\times10^{38}~\text{erg s}^{-1}$ at most feasible effective temperatures.  These limits rule out a SSS system similar to any in the literature with $kT_{\rm eff} > 85$~eV as well as models of accreting, Chandrasekhar-mass WDs with accretion rates $\dot{M}>3\times10^{-8}~M_{\odot}~\text{yr}^{-1}$.
      \item These limits are consistent with WD progenitors that are either low-mass, have low accretion rates, or undergo mass transfer very soon before explosion.  Combined with the limits from other nearby systems, we infer that $<$47\% of SNe~Ia explode from systems involving a stably-accreting Chandrasekhar-mass SSS.
\end{enumerate}

\smallskip\smallskip\smallskip\smallskip
\noindent {\bf ACKNOWLEDGMENTS}
\smallskip
\footnotesize

We thank J. Anais, A. Campillay, and S. Castell\'{o}n for assistance with Swope observations.

The UCSC team is supported in part by NASA grant NNG17PX03C, NSF grant AST--1518052, the Gordon \& Betty Moore Foundation, the Heising-Simons Foundation, and by fellowships from the Alfred P.\ Sloan Foundation and the David and Lucile Packard Foundation to R.J.F.

This work includes data obtained with the Swope Telescope at Las Campanas Observatory, Chile, as part of the Swope Time Domain Key Project (PI Piro, Co-PIs Drout, Foley, Hsiao, Madore, Phillips, and Shappee). This work is based in part on observations collected at the European Organisation for Astronomical Research in the Southern Hemisphere, Chile as part of PESSTO (the Public ESO Spectroscopic Survey for Transient Objects Survey) ESO programmes 188.D-3003, 191.D-0935, and 197.D-1075. This work is based in part on observations made with ESO Telescopes at the La Silla Paranal Observatory under programme ID 099.D-0641 (PI Maguire).

The {\it Hubble Space Telescope} (\hst) is operated by NASA/ESA. Some of our analysis is based on data obtained through programme GO-9427 (PI Harris) from the \hst\ archive operated by STScI.

The scientific results reported in this article are based in part on observations made by the {\it Chandra} X-ray Observatory and published previously in cited articles.  We obtained these data from the {\it Chandra} Data Archive.

\textit{Facilities}: \hst\ (ACS), {\it Chandra} (ACIS), VLT (X-shooter), SOAR (Goodman), NTT (EFOSC), Swope (Direct)

\bibliography{2017ejb}

\end{document}